\documentclass[journal]{IEEEtran}
\pdfoutput=1
\ifCLASSINFOpdf
  \usepackage[pdftex]{graphicx}
\else
\fi
\usepackage{subfigure}
\usepackage{hyphenat}
\usepackage{xcolor}
\usepackage{amsmath}
\usepackage{soul}

\usepackage{epstopdf}
\begin{document}
%
\title{Design and Implementation of Evanescent Mode Waveguide Filters 
Using Dielectrics and Additive Manufacturing Techniques}
%
%
%

\author{Alejandro~Pons-Abenza,~\IEEEmembership{Student~Member,~IEEE,}
        Jos\'e-Mar\'ia~Garc\'ia-Barcel\'o,
        Antonio~Romera-P\'erez,
        Alejandro~Alvarez-Melcon,~\IEEEmembership{Senior~Member,~IEEE,}
        Fernando~Quesada-Pereira,~\IEEEmembership{Member,~IEEE,}
	Juan~Hinojosa-Jim\'enez,~\IEEEmembership{Senior~Member,~IEEE,}
	Marco~Guglielmi,~\IEEEmembership{Fellow,~IEEE,}
	Vicente~Boria,~\IEEEmembership{Fellow,~IEEE,}
        and~Lara~Arche-Andradas,
\thanks{Manuscript received xx xx, 2019; revised xx xx, xxxx.}
\thanks{This work is supported in part by Thales Alenia Space 
(Tres Cantos, Madrid, Spain), and by the grant TEC2016-75934-C4-4-R
in collaboration with TEC2016-75934-C4-1-R
of MEC, Spain.}%
\thanks{A. Pons, J.M. Garc\'ia, A. Romera, A. \'Alvarez, F. Quesada are with Department of Information and Communication Technologies (DTIC), Electromagnetism Applied to Telecommunications research group (GEAT) at Technical University of Cartagena (Murcia, Spain).}
\thanks{J. Hinojosa is with Department of Electronics, Projects and Computer
Technologies at Technical University of Cartagena (Murcia, Spain).}
\thanks{M. Guglielmi and V. Boria are with ITEAM (\textit{Instituto de Telecomunicaciones y Aplicaciones Multimedia}) at Technical University of Valencia (Valencia, Spain).}
\thanks{L. Arche is with Thales Alenia Space Spain (Tres Cantos, Madrid, Spain).}
}

%
%

\markboth{Journal of \LaTeX\ Class Files,~Vol.~XX, No.~X, May~2019}%
{}
%



\maketitle

\begin{abstract}
In this contribution, we describe the design of bandpass filters using
evanescent mode waveguides and dielectric resonators implemented with additive
manufacturing techniques. Two C-band Chebyshev evanescent mode waveguide
filters of order five have
been designed using a low cost commercial dielectric material (ABSplus),
widely used by Fused Deposition Modeling (FDM) 3D printers. 
The housings of the filters have been manufactured using traditional
computer numerical control (CNC) machining techniques. Practical 
manufacturing considerations are also discussed, including 
the integration of dielectric and
metallic parts. We first discuss two breadboards using two 
different resonator geometries. We then demonstrate how 
different transfer functions can be easily implemented by 
changing the 3D printed parts in the same metallic housing.
Breadboards show fractional bandwidths between 3\% and 4.6\% with return
losses better than $RL=18$~dB, and spurious free ranges
of $SFR=1$~GHz. Insertion losses are better than $IL=4.3$~dB.
Even though dielectric losses from the plastic
material are shown to be high, the measured results are quite satisfactory,
thereby clearly showing that this strategy maybe useful for the fast production
of low cost microwave filters implementing complex geometries.
\end{abstract}


%
\IEEEpeerreviewmaketitle

%
%
%
%

\section{Introduction}

\IEEEPARstart{T}{he} design of filters in waveguide 
technology remains one of the most important activities in 
the microwave engineering field \cite{boria07}. 
Waveguide filters are, in fact, 
used in the input/output stages of many systems, for both 
space and ground application, 
where low loss and high-power handling are 
critical requirements \cite{cameron-book}.
In addition, physical constrains such as low volume and small footprint 
are usually required. Furthermore, the most common implementation 
of microwave waveguide filters is based on CNC machining of aluminum. 
Unfortunately, however, CNC machining techniques have limitations 
in terms of manufacturing certain complex geometries \cite{shang17}.

In this context, therefore, emerging Additive Manufacturing (AM) techniques 
are becoming very attractive to produce novel passive filter geometries
and microwave devices \cite{dauria15,peverini17}. 
Using AM techniques complex shapes can, in fact, be very 
easily manufactured \cite{calignano17}.
Using thermoplastic as base material, 
for instance, and using subsequent metallization of the hardware, 
one can dramatically reduce the overall mass of the filter 
structure \cite{montejo15}. In \cite{khan17} the metallization
of the walls
is achieved using an automated deposition process of conductive silver ink.
A similar, although more
elaborated strategy is used in
\cite{chan18}, where air channels are implemented into the walls of the
plastic pieces, which are then filled with a liquid metal. 
It is also interesting to remark that a similar concept was
used in \cite{khan17} to build a waveguide switch. In this case
the switch operation is achieved by pumping in and out the liquid metal
through plastic tubes.
Additionally, smart optimization of the 
shape of traditional implementations can significantly reduce the
Insertion Losses ($IL$) of the filters \cite{lorente09}. 
Other authors have tried direct geometrical modifications 
on the cavity resonator itself, using curvilinear shapes, 
leading to new resonators with significant improvements in 
terms of unloaded quality factors ($Q_U$). Interesting works
in this direction are the spherical resonators proposed in
\cite{Guo16,guo15}, or the use of complex curvilinear shapes
as proposed in \cite{booth17}.

The implementation of dielectric resonators
has also been recently explored in the context of AM. 
A list of thermoplastics and ceramics showing 
good dielectric properties can, in fact, be found for 
many FDM or Stereo-lithography (SLA) commercial 3D-printers 
\cite{salonitis03}, \cite{ngoc17}.
Various strategies have also been discussed to incorporate AM dielectrics 
into a waveguide
filter structure. An interesting example can be found in \cite{carceller17}, 
where a TM-mode dielectric
resonator based on alumina with 
$\tan \delta < 0.002$ and $\epsilon_r \approx 9.9$
is used to directly print a second order filter.  After
the hardware has been printed,
the external faces are metallized to build the lateral walls of the structure. 
The filter shows
good performance in terms of losses, but with
slight frequency deviations due to the lack of accuracy of the AM process. 
An alternative solution is to design
the dielectric resonators from a single piece of plastic,
which is then fastened in the filter using the metallic parts of the 
coupling windows.
This strategy is demonstrated in \cite{marchives14}, where the dielectric 
resonators are printed from a single plastic bar,
following an in-line topology, or in \cite{perigaud18}, where a more 
complex connection mechanism between resonators is proposed.

\begin{figure}
\centering
\includegraphics[width=\columnwidth]{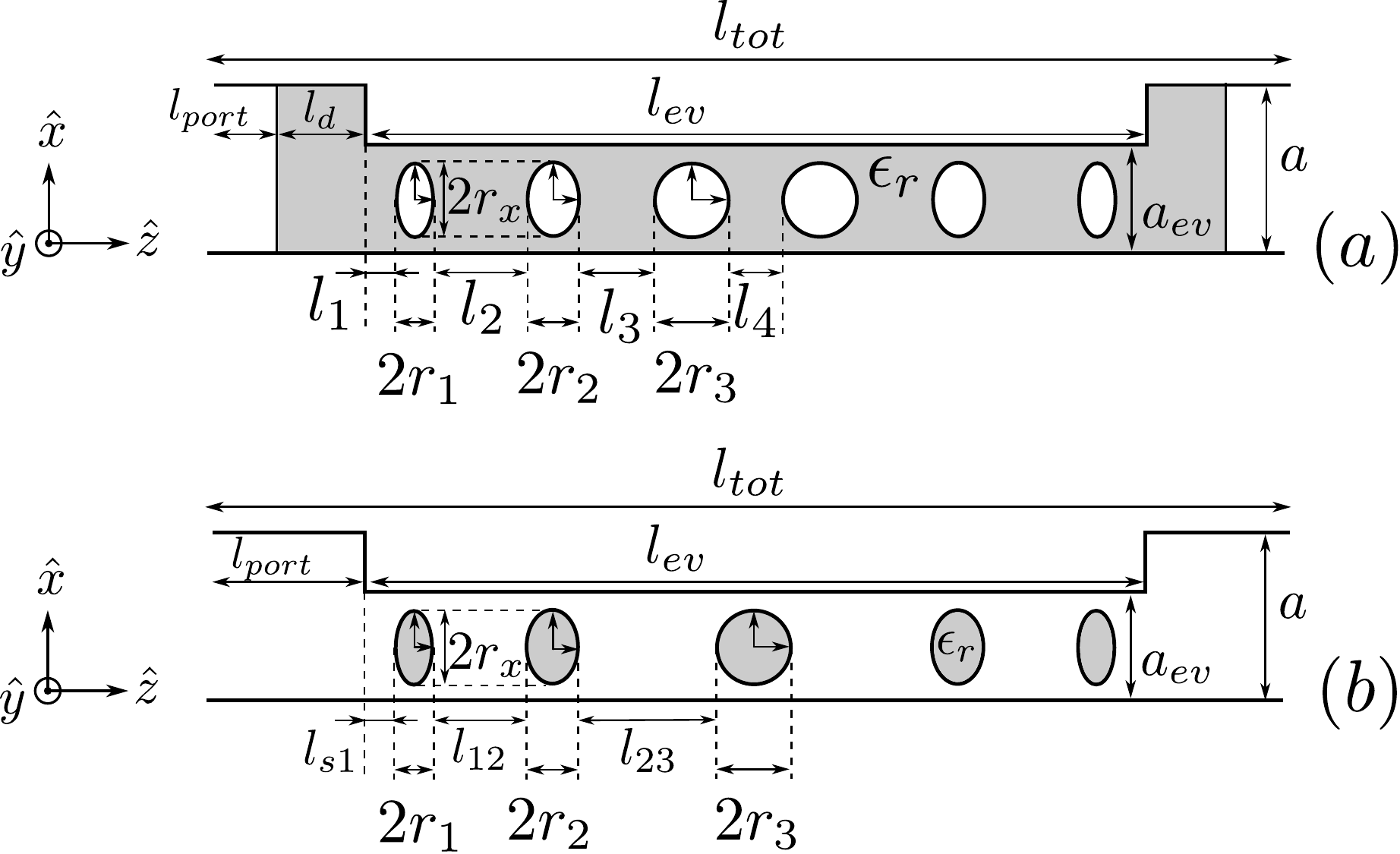}
\caption{2D sketches of the evanescent mode filters proposed in this paper. 
(a) Design with air holes inside a dielectric block filling a
waveguide section with reduced width (air-holes filter).
(b) Design with dielectric posts inside an otherwise
empty evanescent waveguide section ($\epsilon_r$-posts filter).}
\label{fig:evanescent_sketches}
\end{figure}

In this context therefore, we contribute to the 
state-of-the-art of AM manufacturing of waveguide filters by showing how the 
two evanescent mode filter designs shown in 
Fig.~\ref{fig:evanescent_sketches} can be manufactured using 
3D printing techniques. 
The objective of this paper is to compare 
two different solutions for evanescent mode filters  
that can be rapidly and inexpensively manufactured with 
3D-printing using commercial AM dielectric materials. 
The filter structures discussed in 
this paper combine AM parts for the dielectric 
elements, with conventional aluminum CNC machining, 
thus leading to a hybrid manufacturing strategy. 
Following this strategy, the metallic parts of the structure
are kept simple, while complex geometries may be easily manufactured
for the dielectric elements.

The dielectric elements proposed in this work are implemented with a 
low cost commercial ABSplus plastic from Stratasys, with
relative dielectric constant $\epsilon_r=3.55$ \cite{ABSplus}. 
This material has recently become widely available in low cost 3D printers.
The first solution, shown in Fig.~\ref{fig:evanescent_sketches}(a),
consists of a filter with a reduced width,
dielectric-filled rectangular waveguide, where 
inter resonator couplings are implemented 
with elliptical air holes (subsequently referred to as the air-hole filter).
The complementary approach is followed in the second design,
where several elliptic dielectric post resonators are placed
inside an empty evanescent rectangular waveguide section,
as shown in Fig.~\ref{fig:evanescent_sketches}(b) 
(subsequently referred to as the $\epsilon_r$-post filter). 

The two structures proposed
in this work have been manufactured with a hybrid strategy. Namely,
the 3D printed plastic pieces have 
been manufactured using an FDM 3D printer, while the 
aluminum housings have been manufactured
using standard CNC machining. This
strategy allows for a fast and low cost implementation of the filters,
although at the expense of somewhat higher insertion losses.
An additional benefit of this hybrid manufacturing strategy is also
demonstrated in this paper. Namely, several
filters with different transfer functions can 
be easily implemented using the same external aluminum housing. 
What is needed is just the design of new dielectric 
parts, which are then manufactured
with the 3D printer. Following this approach, only one \textit{common}
external housing is manufactured with CNC machining techniques,
while high flexibility can be obtained by producing
several low cost plastic pieces to implement different transfer
functions. 

Overall, the measured performance of the manufactured
prototypes is good, thereby fully
confirming the practical feasibility of the
new concepts proposed in the paper.
\section{Filters design}
The filters used in this work to demonstrate the 
hybrid manufacturing strategy are two 5-pole Chebyshev 
filters with $20$~dB return 
loss ($RL$). The filters have been designed at C-band with a center 
frequency $f_c = 3.68$~GHz and a bandwidth $BW = 120$~MHz 
(fractional bandwidth $FBW \approx 3.2\%$). The filters are 
designed using an in-line topology, since no cross-couplings 
will be considered for simplicity. It is important to note, however, 
that cross coupled structures can also be easily implemented. 
The coupling matrix, 
which is the same for both filters, has all elements equal
to zero, except for couplings along the in-line channel ($M_{i,i+1}$)
\cite{cameron-book}
\begin{equation}
\label{eqn:couplingMatrix}
\begin{split}
M_{S1} = M_{5L} = 1.0137, \, M_{12} = M_{45} = 0.8653, \\
M_{23} = M_{34} = 0.6357.
\end{split}
\end{equation}

2D sketches of the proposed filters are presented in 
Fig.~\ref{fig:evanescent_sketches}. A standard WR-229 waveguide has 
been used for the input/output waveguide ports 
($a = 58.17$~mm, $b = 29.083$~mm). 
The first filter, shown
in Fig.~\ref{fig:evanescent_sketches}(a), consists of a 
dielectric piece placed inside a reduced width
waveguide section, with a series of air holes. This 
filter uses the fundamental TE$_{10}$ mode, 
which can propagate in the dielectric filled waveguide section, 
while the air holes, where the fundamental mode is below cut-off, 
are used to implement the coupling elements between resonators. The width
of the waveguide section is reduced to avoid the 
propagation of higher order modes, when it is filled with the
dielectric material.
The physical separation between air holes 
[$l_i$ in Fig.~\ref{fig:evanescent_sketches}(a)] controls the
resonant frequency of each resonator. The air holes are 
implemented as elliptical cylinders, and their
dimensions [$r_i$ according to 
Fig.~\ref{fig:evanescent_sketches}(a)] are used to control 
the coupling level between adjacent resonators. Consequently, 
for the design of our 5th-order filter, a total of six 
air holes are required.  With this arrangement,
the input/output couplings are controlled with the sizes of
the first and last air holes. However, the step discontinuity
needed for implementing the reduced width waveguide section
will also affect this coupling. 

As it can be observed in
Fig.~\ref{fig:evanescent_sketches}(a), the dielectric piece
is longer than the waveguide section ($l_{ev}$). The
extra length ($l_d$) will facilitate the alignment of the
dielectric piece with the external metallic housing.
We also verified that this extra dielectric section helps to
increase the input/output coupling levels of the filter.
However, the extra length ($l_d$) cannot be very large, as this
area may become resonant due the formation of an
air-dielectric discontinuity. As it will be shown
later in this paper, this suggests that, in practice,
a trade-off is needed to adjust the length $l_d$.

The second solution proposed in this paper
is shown in Fig.~\ref{fig:evanescent_sketches}(b). 
In this case, instead of filling the whole waveguide 
with a dielectric material, only several cylindrical dielectric posts 
are placed inside an empty waveguide section. With this
arrangement the fundamental TE$_{10}$ mode is below cut-off
in the evanescent waveguide section, and the dielectric
posts become resonant. Consequently,
the dimensions of these posts [$r_i$ according 
to Fig.~\ref{fig:evanescent_sketches}(b)] control the resonant 
frequency of each resonator, while the physical distances between 
adjacent posts [$l_{i,i+1}$ in Fig.~\ref{fig:evanescent_sketches}(b)]
control the coupling level by proximity. For our 5th-order 
filter, a total of 5 dielectric posts are required. 
Intuitively, this solution may be more compact, since less posts 
(as compared to air-holes)
are required to implement the same frequency response. 
For this solution,
the input/output couplings are implemented with
the distance between the ports and the first (or last) dielectric
resonator [$l_{s1}$ in Fig.~\ref{fig:evanescent_sketches}(b)].
However, the maximum coupling that can be achieved is
strongly determined by the step junction
needed to form the evanescent waveguide section.

{It is important to note that in both structures 
shown in Fig.~\ref{fig:evanescent_sketches},
the cylinders are of
elliptical shape, and we have fixed the radius along the
$x$-axis to $r_x = 14$~mm. This assures
a $1$~mm gap between the waveguide side walls and each post,
for a width of the evanescent waveguide section of $a_{av}=30$~mm.}

\begin{figure}
\centering
\includegraphics[width=\columnwidth]{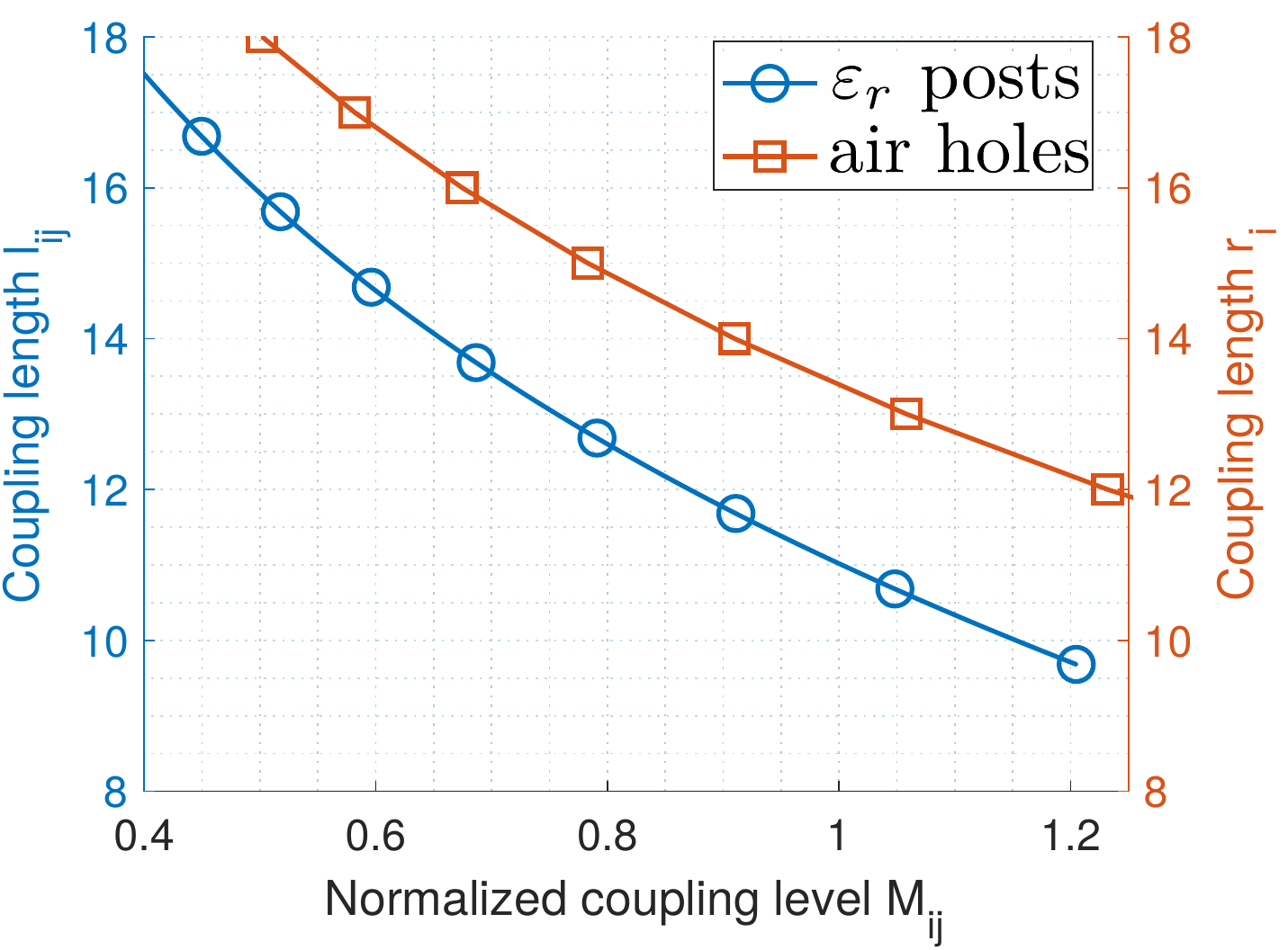}
\caption{Normalized inter-resonator 
	coupling curves for the two proposed 
	configurations. Legend refers to 
	Fig.~\ref{fig:evanescent_sketches}(a) (air-holes)
	and to Fig.~\ref{fig:evanescent_sketches}(b) ($\epsilon_r$-posts).
	For this test ($a_{ev}=30$~mm).}
\label{fig:mij_curves}
\end{figure}

For the design process, a variation of the classical even/odd 
analysis technique has been used \cite{pons18}. 
This approach is based on calculating the even-odd resonant
frequencies ($f_e$, $f_o$) of two symmetric coupled 
resonators working at the 
center frequency ($f_c$). The frequency scaled coupling 
coefficient can then directly be computed from these two frequencies as
\cite{cameron-book,matthaei80}
\begin{equation}
\label{eqn:ksc}
k_{sc} = \frac{f_e^2-f_o^2}{f_e^2+f_o^2},
\end{equation}
and the normalized coupling coefficient, which is directly 
related to the coupling matrix values given in
\eqref{eqn:couplingMatrix}, can be calculated as 
\cite{cameron-book,matthaei80}
\begin{equation}
\label{eqn:Mij}
k_{norm} = M_{ij} = \frac{f_c}{BW} \, k_{sc}.
\end{equation} 
The normalized coupling values ($M_{ij}$), obtained 
for the two types of resonators we propose,
are presented in Fig.~\ref{fig:mij_curves}.
The same figure is employed to present the results obtained for
the two configurations discussed in this paper, by using two independent
vertical axes.
For the configuration shown in Fig.~\ref{fig:evanescent_sketches}(a),
the coupling is given as a function of the air-hole semi-axis ($r_{i}$),
while the length ($l_{ij}$) is used for the second configuration
shown in Fig.~\ref{fig:evanescent_sketches}(b).

The levels computed for $M_{ij}$ are suitable in both cases 
to obtain the filters with the initial specifications. 
It can be observed that, to implement an equal coupling level, the
configuration with dielectric posts will in general be smaller
than  the one with air-holes ($l_{ij}<r_i$). This indicates that,
for equal responses, the solution with dielectric posts 
will result in a more compact structure.

\begin{figure}
\centering
\includegraphics[width=\columnwidth]{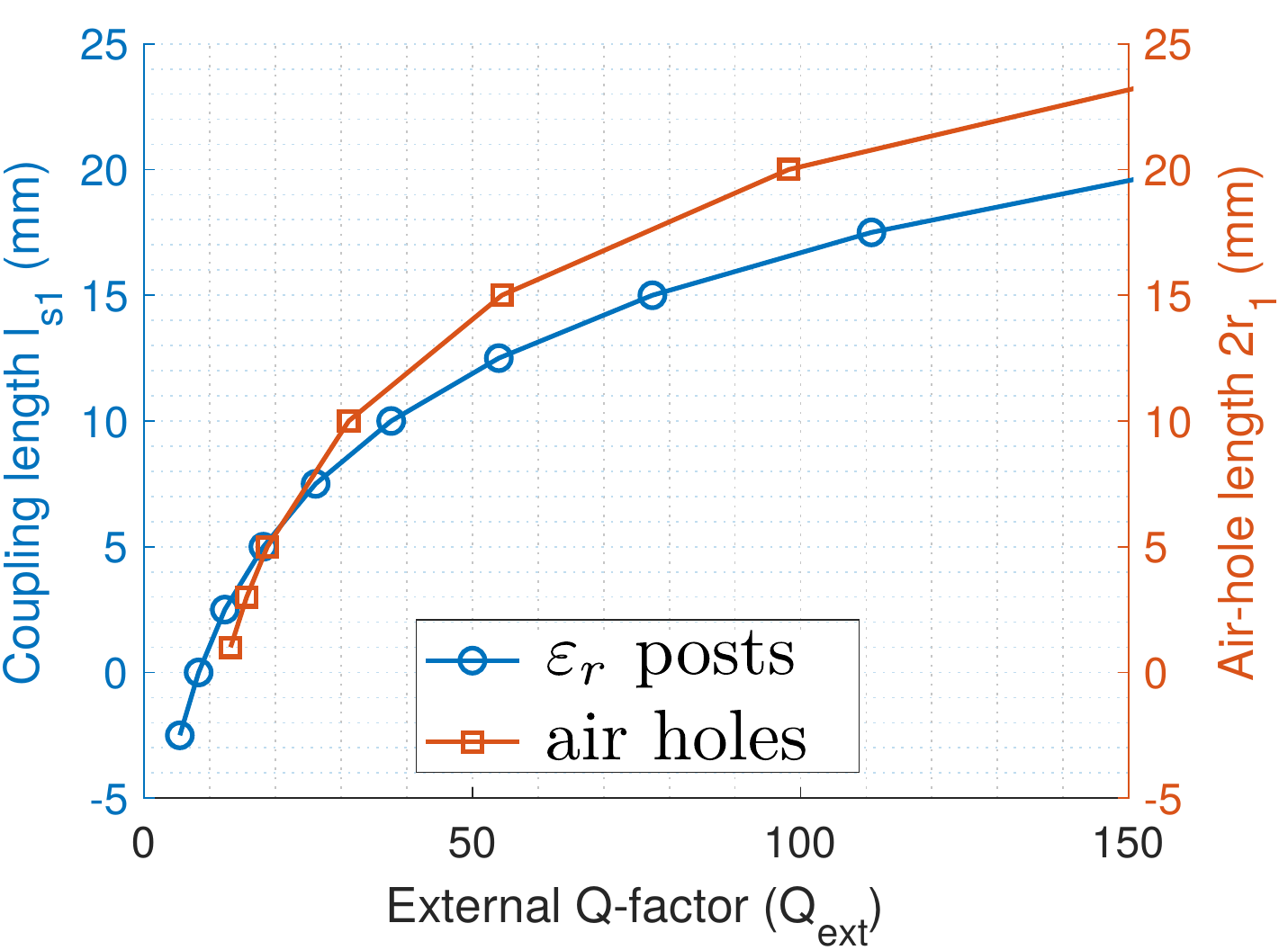}
\caption{External quality factor ($Q_{ext}$) for the two
	proposed evanescent filters. The legend refers to
	Fig.~\ref{fig:evanescent_sketches}(a) (air-holes)
	and to Fig.~\ref{fig:evanescent_sketches}(b) ($\epsilon_r$-posts).
	For this test we have fixed: $a_{ev}=30$~mm, $l_d=7$~mm,
	$l_1=0$~mm.}
\label{fig:qext_curves}
\end{figure}

It is well known that
for the input/output coupling, a different 
procedure must be followed. This procedure consists of 
adjusting the external quality factor ($Q_{ext}$)
of a single resonator to the
value calculated from the coupling matrix, namely
\cite{cameron-book,matthaei80}
\begin{equation}
\label{eq:qext}
  Q_{ext} = \dfrac{f_c}{M_{S1}^2 \, BW}.
\end{equation}
A doubly-terminated 
resonator structure can be used for this purpose. This 
will allow the computation of the $Q_{ext}$
as a function of the physical parameter of the filter related to the
input/output couplings. For our two configurations,
the key parameters are the elliptical semi-axis of the first air-hole
($r_1$) [air-hole filter, Fig.~\ref{fig:evanescent_sketches}(a)] 
and the distance to the first resonator ($l_{s1}$)
[dielectric post filter, Fig.~\ref{fig:evanescent_sketches}(b)].
The corresponding results for each structure are shown 
in Fig.~\ref{fig:qext_curves}. Similarly to 
Fig.~\ref{fig:mij_curves}, two different ordinate axes have been used 
to show both results on the same plot.

As we can see from the $Q_{ext}$ curves shown in Fig.~\ref{fig:qext_curves}, 
the slopes are similar. However,
the structure with dielectric posts can achieve lower 
$Q_{ext}$ values. Therefore, this structure
is more suitable to implement filters with wider passbands as compared to the
filter with air-holes.
This behavior is enhanced by the fact that the coupling
distance $l_{s1}$ in Fig.~\ref{fig:evanescent_sketches}(b) 
can be made slightly negative, meaning
that the first and last resonators are slightly introduced in
the port section [$l_{port}$ of Fig.~\ref{fig:evanescent_sketches}(b)].

Another interesting conclusion can be pointed out from the
plot shown in Fig.~\ref{fig:qext_curves}.
For narrowband filters, where larger $Q_{ext}$ values are
needed, the filter with dielectric posts has the potential
to be more compact. As shown in the plot, large
$Q_{ext}$ values require smaller
coupling distances ($l_{s1}$) for the dielectric
posts filter, as compared to the diameter ($2\,r_1$)
of the air-holes filter ($l_{s1}<2\,r_1$). For instance, to
synthesize $Q_{ext} = 100$, a coupling length
$l_{s1} = 16$~mm is needed, while the axis of
the air-hole needs to be larger ($2\,r_{1} = 20$~mm).
This confirms the behavior already observed for the
inter-resonator couplings in Fig.~\ref{fig:mij_curves}.
For the same in-band response, more compact structures can
be obtained with the configuration based on dielectric posts.

\begin{figure}
\centering
\includegraphics[width=\columnwidth]{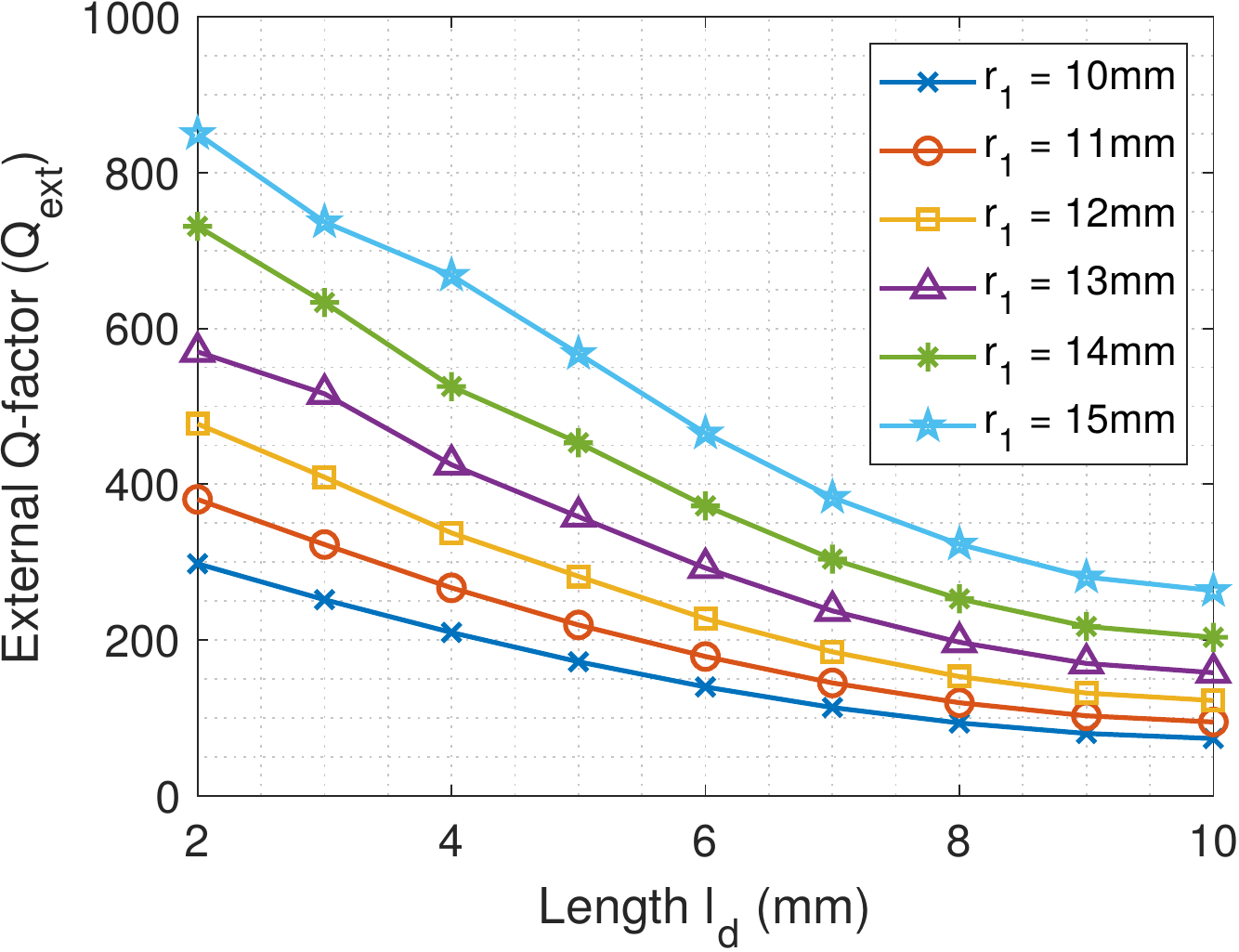}
\caption{$Q_{ext}$ values for the air-holes filter as a 
         function of the extra length ($l_d$),
         for various values of the input semi-axis ($r_1$). For this test
         we have fixed: $a_{ev}=30$~mm, $l_1=0$~mm.}
\label{fig:qext_airhole_ld}
\end{figure}

As already mentioned, the input coupling
of the air-hole filter is also affected
by the extra length ($l_d$) used to align the dielectric
piece with the external housing. In order to study this
effect, we present in  Fig.~\ref{fig:qext_airhole_ld}
the $Q_{ext}$ values for this filter as a function
of the extra length ($l_d$) for various values
of the input air-hole semi-axis ($r_1$).
As we can see, the $Q_{ext}$ value decreases when
the length $l_d$ is increased. This indicates that 
$l_d$ can be used to increase the input
coupling of the filter, if wideband responses are needed.
On the other hand, the plot shows that 
the slope is greater when the semi-axis of the
air-hole ($r_1$) is larger. Therefore, the sensitivity of the input
coupling with the length ($l_d$) will be stronger for narrowband
responses. The plot also indicates that the
effect tends to saturate for lengths $l_d>8$~mm. Further increase
of this length does not translate in substantially higher
input coupling.
In addition, we have verified that it is not convenient
to take very large values for the length $l_d$. This is because the
junction may become resonant in the area between 
the waveguide step and the air-dielectric discontinuity.
Taking into consideration all the practical aspects discussed so far,
and the behavior
shown in Fig.~\ref{fig:qext_airhole_ld}, an optimum value
of $l_d=7$~mm has been selected for the designs presented in this paper.

\begin{figure}
\centering
\includegraphics[width=\columnwidth]{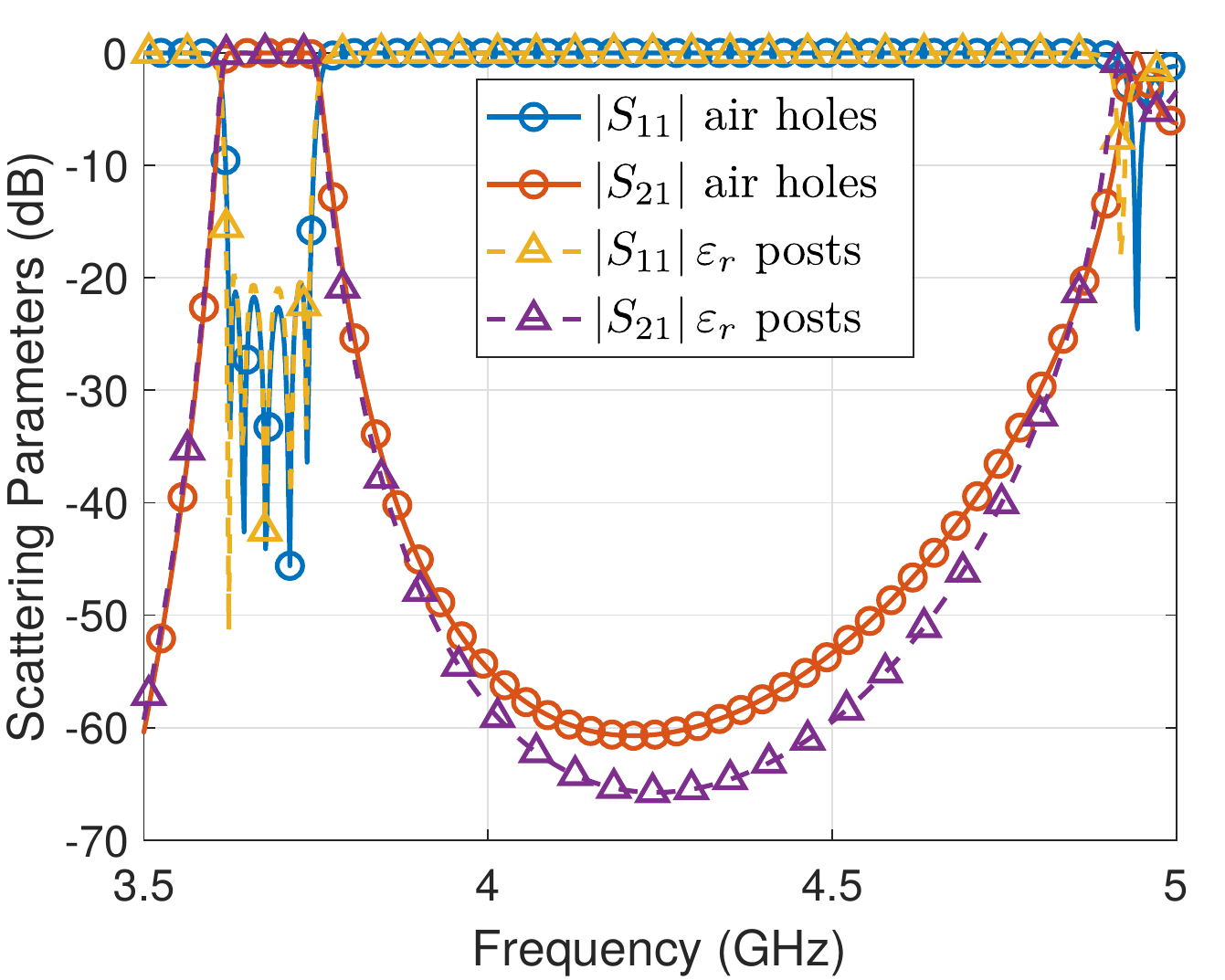}
\caption{Frequency responses obtained for the two designed filters.
        The legend refers to
        Fig.~\ref{fig:evanescent_sketches}(a) (air-holes)
        and to Fig.~\ref{fig:evanescent_sketches}(b) ($\epsilon_r$-posts).} 
\label{fig:filter_ideal_responses}
\end{figure}

Using the data shown in Fig.~\ref{fig:mij_curves}
and Fig.~\ref{fig:qext_curves}, the initial dimensions
for the two filters can now be easily obtained. 
The final \textit{optimized} filter dimensions are shown
in Table~\ref{tab:evanescente_aire} and
Table~\ref{tab:evanescente_postes}, respectively.
\begin{table}
\centering
\caption{Final dimensions of the designed air-holes filter
	according to Fig.~\ref{fig:evanescent_sketches}(a).}
\begin{tabular}{c|c||c|c||c|c} 
	\textit{Var.} & Value (mm) & \textit{Var.} & Value (mm) & 
	\textit{Var.} & Value(mm)\\
\hline
$l_{port}$ & $10$ & $l_{1}$ & $0$ & $r_1$ & $4.859$ \\ 
$l_{d}$ & $7$ & $l_{2}$ & $22.11$ & $r_2$ & $14.54$ \\ 
$l_{ev}$& $258.847$ & $l_{3}$ & $23.23$ & $r_3$ & $16.86$ \\ 
$l_{tot}$& $292.847$ & $l_{4}$ & $23.13$ & $r_x$ & $14$ \\ 
\end{tabular}
\label{tab:evanescente_aire}
\end{table}
\begin{table}
\centering
\caption{Final dimensions of the designed dielectric posts filter
	according to Fig.~\ref{fig:evanescent_sketches}(b).}.
\begin{tabular}{c|c||c|c||c|c} 
	\textit{Var.} & Value (mm) & \textit{Var.} & Value (mm) & 
	\textit{Var.} & Value (mm)\\
\hline
$l_{port}$ & $15$ & $l_{s1}$ & $-0.439$ & $r_1$ & $13.619$ \\ 
$l_{tot}$ & $275.016$ & $l_{12}$ & $24.6$ & $r_2$ & $14.208$ \\ 
$l_{ev}$& $245.016$ & $l_{23}$ & $28.474$ & $r_3$ & $14.219$ \\ 
\end{tabular}
\label{tab:evanescente_postes}
\end{table}
The wide band responses for both filters are presented
in Fig.~\ref{fig:filter_ideal_responses}. 
The results shown have been obtained with the full-wave simulator
HFSS~\cite{hfss}.
As we can see, the responses of the two filters are
very similar, both inside and outside the passband. 
The out of band responses indicate
that the spurious free ranges (SFR) of both filters are also very similar.
In particular, the first spurious band appears at
$f_{sp} \approx 4.8$~GHz, giving $SFR \approx 1$~GHz. 

Except for the insertion losses ($IL$), that will be discussed in the
next section, the results obtained
indicate that the performance of both filters is essentially the same one,
at least from the electrical point of view. Mechanically, however,
there are some important differences. 
First, the dimensions
of Table~\ref{tab:evanescente_aire} indicate that the total
length of the filter with air-holes [Fig.~\ref{fig:evanescent_sketches}(a)]
is $l_{tot} = 292.847$~mm. For the second filter
[Fig.~\ref{fig:evanescent_sketches}(b)], the data of
Table~\ref{tab:evanescente_postes} indicate a total
length of $l_{tot} = 275.016$~mm. This confirms that,
as previously discussed, the filter based on dielectric posts is
more compact than the filter based on air-holes. This can be
a convenient feature for practical applications, where reduction in
footprint is important.
\section{Experimental results}
The two filters designed in the previous section have been
implemented using a hybrid manufacturing approach.
AM has been used for the dielectric parts, 
while the metallic housings have been manufactured 
using standard CNC milling. The metallic parts
have been manufactured using aluminum, while the
dielectric parts have been printed using ABSplus with a 3D-printer,
model Dimension 1200-bst \cite{3Dprinter}.

\begin{figure}
\centering
\includegraphics[width=\columnwidth]{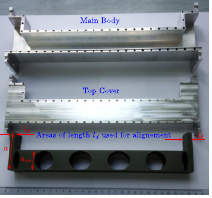}
\caption{Pieces of the manufactured air-holes filter, showing the two
	extended areas of length ($l_d$) used for alignment of the dielectric 
	block inside the main body.}
\label{fig:photo_air_post_filter}
\end{figure}

The pieces manufactured for the air-hole
filter are shown in Fig.~\ref{fig:photo_air_post_filter}.
As we can see, the housing is manufactured in two pieces, namely, a main body
and a top cover. This geometry is very simple, so a standard CNC milling
process can be used. The dielectric piece is manufactured with
the 3D printer in one single block, including all 
resonators and couplings. As a result,
the alignment of the dielectric piece inside the
metallic body is extremely simple. The alignment is also 
facilitated by the use of the
two extended areas of length $l_d=7$~mm at both extremes
of the piece, as shown in Fig.~\ref{fig:photo_air_post_filter}.

\begin{figure}
\centering
\includegraphics[width=\columnwidth]{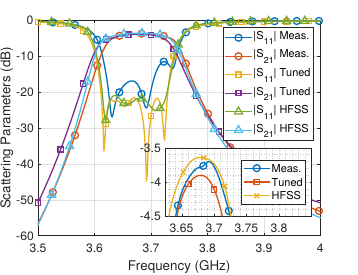}
\caption{S-Parameter comparison between measured results directly
	obtained after manufacturing (circle symbols), measured results
	after tuning (square symbols) and simulated results obtained
	with HFSS after introducing losses (triangle and cross symbols)
	($\tan \delta=0.0053$, $\sigma=3.8\cdot10^7$~S/m). Inset in the
	figure shows details of the measured and simulated
	minimum insertion
	losses within the passband.} 
\label{fig:air_post_measurements}
\end{figure}

The measured results, directly obtained from the 
prototype manufactured, are shown in Fig.~\ref{fig:air_post_measurements}
using circle symbols. They are compared with the results
obtained with HFSS when losses are included in the simulations
(triangle and cross symbols). As we can see, the response obtained
directly after manufacturing is in reasonably good agreement 
with the simulations performed with HFSS including losses. 

To include losses in the simulations, we have used for 
the ABSplus a loss tangent value equal to $\tan \delta = 0.0053$. This value 
corresponds to the
largest value specified by the manufacturer 
($0.0046 < \tan \delta < 0.0053$) \cite{ABSplus}. 
In addition, for the external aluminum housing,
the conductivity value used is
($\sigma=3.8\cdot10^7\,$S/m). However, we have
verified that the losses due to the finite conductivity of the  
external housing have
very little influence on the overall insertion loss performance
of the filter. This study confirms that the important
contribution of losses is due to the loss tangent of the
ABSplus. As observed in the inset
of Fig.~\ref{fig:air_post_measurements}, the minimum insertion
loss measured inside the passband is $IL=3.7$~dB.

Apart from the insertion losses, the other measured characteristics
are: center frequency 
$f_c=3.671$~GHz, and bandwidth $BW=117$~MHz. 
Overall, the measured response shows good agreement in terms 
of central frequency and filter bandwidth, as compared to the 
ideal design specifications. This is a clear indication that the
relative permittivity of the ABSplus block is very close
to the nominal value reported by the manufacturer 
($\epsilon_r = 3.55$) \cite{ABSplus}.
On the contrary, the measured minimum return loss within the
passband is worse ($RL=12$~dB) than the expected value.
In addition, we can clearly see that a number of reflection poles
are not clearly visible. This means that the filter is not correctly tuned.
In general, this behavior is normal with low accuracy manufacturing processes.
The return loss is indeed the most sensitive parameter
in the response of a microwave filter. 

\begin{figure}
\centering
\includegraphics[width=\columnwidth]{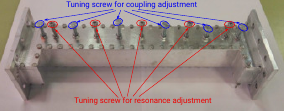}
\caption{Photograph of the manufactured filter based on
	air holes (Fig.~\ref{fig:evanescent_sketches}(a)), 
	after assembling. Details of the tuning screws used to compensate
	for manufacturing errors are also shown.}
\label{fig:photo_air_post_screws}
\end{figure}

However, the detuning observed in the return
loss response can be corrected by introducing tuning screws in
the manufactured prototype, as shown in Fig.~\ref{fig:photo_air_post_screws}.
In this case, standard $M3$ tuning screws were used.
As we can see in the figure,
a total of 11 tuning screws have been used for the fine
tuning of the filter. Six screws
are located in the center of the air holes, in order to
adjust the couplings. Five more screws
are placed in the middle of the resonators (between two contiguous air holes), 
to allow for the adjustment of the resonant frequencies of the individual
resonators. For the installation of these last screws, the dielectric
block has been perforated at the appropriate locations.
In this context, it is interesting to note that 
the perforations will shift the resonances to
higher frequencies. The presence of the screws in the resonators, 
however, will lower the resonant frequencies,
thus compensating for the initial frequency shifts. This is the reason why the
final prototype shown in Fig.~\ref{fig:photo_air_post_screws}
can be tuned,  keeping the same center frequency as in the
original structure without tuning screws. 

The response of the hardware after tuning is also included in
Fig.~\ref{fig:air_post_measurements} (with square symbols).
Results show that the tuning process was
indeed effective. In particular, the filter has the same center frequency and
bandwidth as for the initial specifications. Furthermore, 
an almost equiripple return loss response has been obtained with a
$RL=21$~dB. However, the improvements in the passband response come at the
expense of a limited increase
in the minimum insertion loss of the filter within 
the passband, which now increases to $IL=3.9$~dB, as shown
in the inset of Fig.~\ref{fig:air_post_measurements}.
Overall, a good filter performance has been demonstrated
with the manufactured prototype. Also, the tuning
process has shown to be effective in compensating the
manufacturing errors introduced by the 3D printing technique.

\begin{figure}
\centering
\includegraphics[width=\columnwidth]{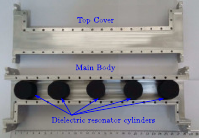}
\caption{Pieces of the manufactured filter with dielectric post
	resonators. The five dielectric elliptic cylinders are shown already
	mounted and aligned inside the empty evanescent waveguide section.}
\label{fig:photo_er_post_filter}
\end{figure}

\begin{figure}
\centering
\includegraphics[width=\columnwidth]{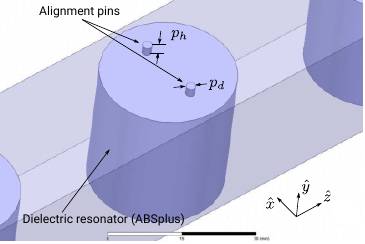}
\caption{Details of a dielectric cylinder resonator including
	the two pins incorporated for alignment inside the
	empty evanescent waveguide section. The alignment pins
	are place along the $\hat{x}$-axis.}
\label{fig:pins_3d}
\end{figure}

The second filter, based on dielectric posts, 
as shown in Fig.~\ref{fig:evanescent_sketches}(b), has been manufactured
using a similar strategy. Again, the metallic housing is
manufactured from aluminum in two pieces, with a main body and a top cover,
as shown in Fig.~\ref{fig:photo_er_post_filter}.
In this case, an additional practical challenge
is the alignment of the five elliptical dielectric resonators inside the
empty evanescent waveguide section.
To meet this challenge, in the 3D printing of 
the five dielectric resonators we have  included two alignment pins
of height $p_h$ and diameter $p_d$, respectively, as shown
in Fig.~\ref{fig:pins_3d}.
Two holes per resonator are then drilled in the bottom wall of the
main body, with slightly greater diameter
than the alignment pins. The dielectric 
cylinders are then simply placed inside the metallic housing
by introducing the pins in the corresponding holes.
For practical considerations related to the
the 3D printing process,
the height and diameter of the pins were selected
to be $p_h = 2$~mm and $p_d = 2$~mm, respectively. With these dimensions,
we have verified with EM simulations that
the pins do not have any 
significant influence on the electrical performance
of the structure.

\begin{figure}
\centering
\includegraphics[width=\columnwidth]{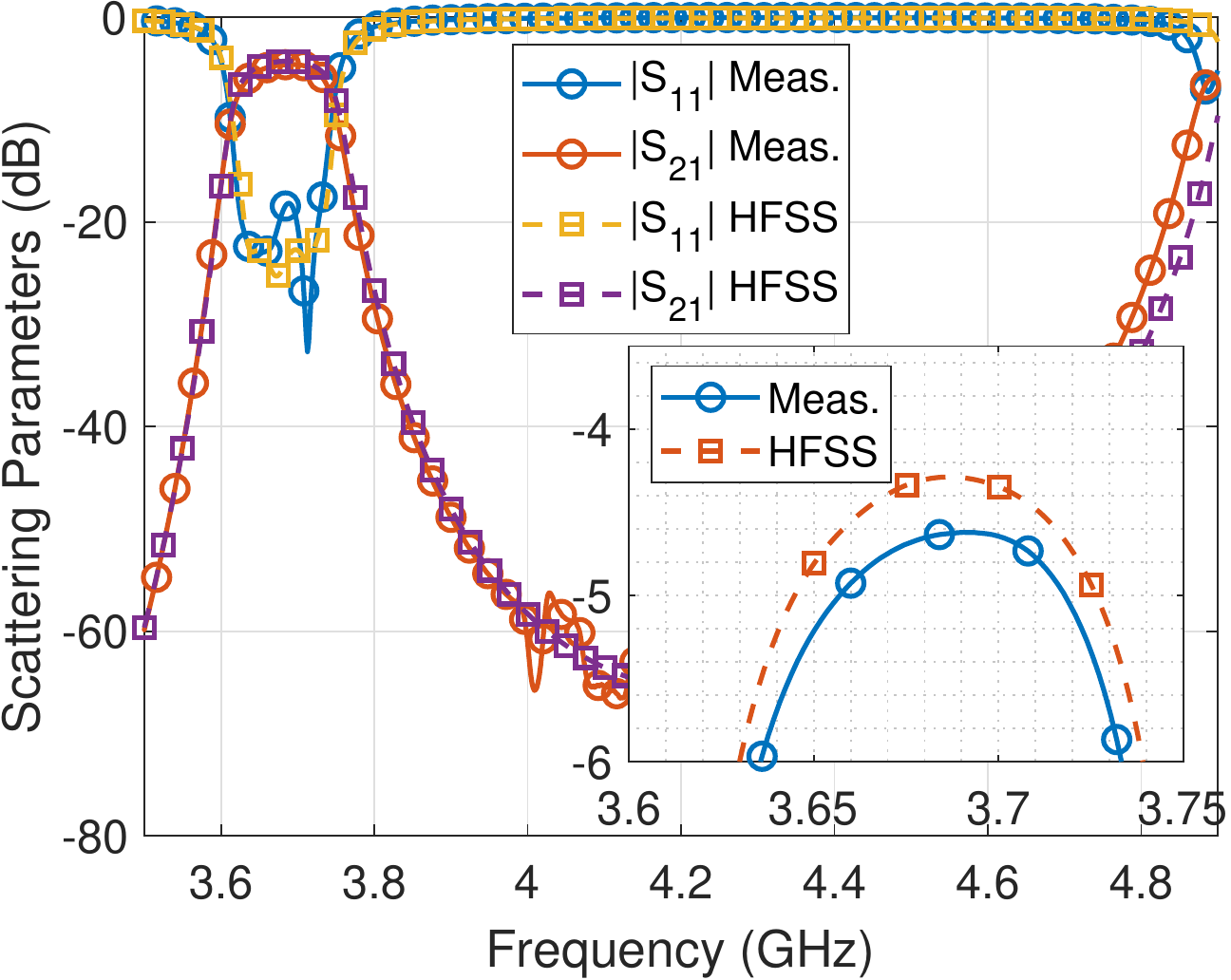}
\caption{Comparison between the measured and simulated S-parameters,
	for the dielectric posts filter. Simulations include losses
	in the dielectric resonators ($\tan \delta=0.0053$) and
	finite conductivity in the housing ($\sigma=3.8\cdot10^7$~S/m).}
\label{fig:er_post_measurements}
\end{figure}

After assembling the dielectric cylinders, 
the measured scattering parameters of the
filter are shown in Fig.~\ref{fig:er_post_measurements}.
Comparison with respect to the simulated response is also
shown, after including losses in the analysis of the structure.
As we can see, the measured and simulated results
show excellent agreement for both center frequency and bandwidth. 
The $RL$ performance inside the passband without tuning is better in this case, 
with a minimum value of $RL = 18$~dB. The measured bandwidth
at a constant return losses of 20~dB is $104$~MHz. In addition, 
the measured minimum
insertion loss within the passband is ($IL=4.3$~dB)
(see the inset of Fig.~\ref{fig:er_post_measurements}). 
This insertion loss is only $0.3$~dB higher than the
simulated response using $\tan \delta = 0.0053$ to
model the ABSplus plastic, and the conductivity of
aluminum for the housing. As it happened 
with the previous design, the loss tangent
of the ABSplus plastic seems to be in the upper limit
reported by the manufacturer 
($0.0046 < \tan \delta < 0.0053$) \cite{ABSplus}.
Since in this case the agreement between measurements and simulations is
good, we did not use any further post-manufacturing tuning 
mechanism.

\begin{figure}
\centering
\includegraphics[width=\columnwidth]{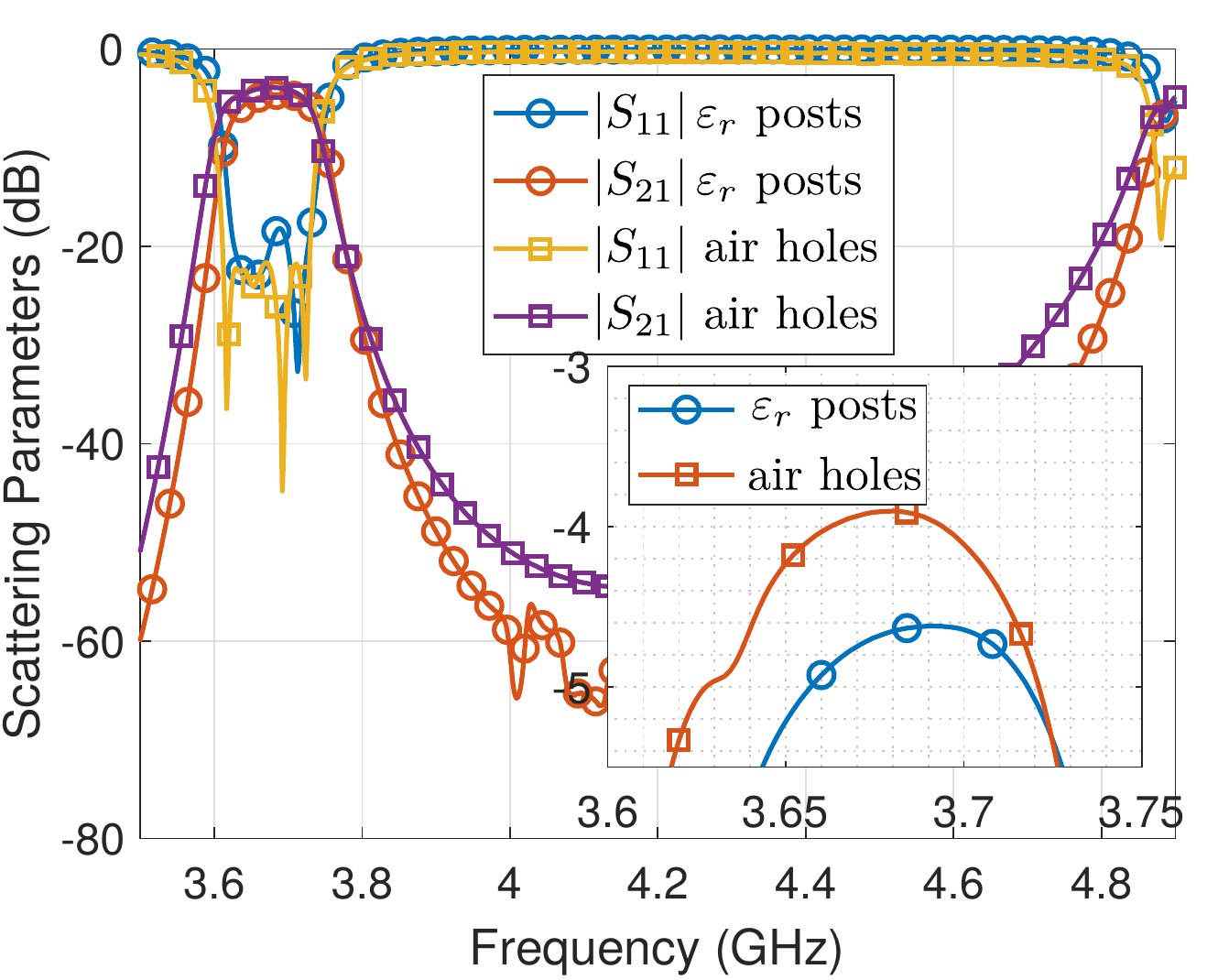}
\caption{{Comparison between measured results of 
	the two manufactured filters with
	$BW = 120\,$MHz.}
	Legend refers to 
        Fig.~\ref{fig:evanescent_sketches}(a) (air-holes)
        and to Fig.~\ref{fig:evanescent_sketches}(b) ($\epsilon_r$-posts). 
	The tuned response obtained with screws is used 
	for comparison in the case of
	the air-holes filter.}
\label{fig:air_vs_er_measurements}
\end{figure}

Finally, in Fig.~\ref{fig:air_vs_er_measurements} 
we include a comparison between
the measured responses obtained for the two filters that we have manufactured.
The response of the air-hole filter used for this comparison
is the one obtained after tuning 
with screws (see Fig.~\ref{fig:photo_air_post_screws}).
Details of insertion losses can be observed in the
inset of the figure. As shown, the filter based on air-holes has
lower insertion losses ($IL=3.9$~dB) as compared to the 
dielectric post filter ($IL=4.3$~dB). 
This can be explained by the larger dimension of this filter
($l_{tot}=292.847$~mm) as compared to the dielectric 
post filter ($l_{tot}=275.016$~mm). The extracted $Q_U$ values from
both filters are $Q_U=305.75$ and $Q_U=303.57$, respectively.
This indicates that the solution based on air-holes
leads to resonators of slightly higher $Q_U$, although at the
expense of somewhat larger dimensions.
Except for the insertion losses,
the performances of the two filters are very similar, both inside and outside
of the passband. In particular, both filters
exhibit very similar measured $SFR$ behaviors, with a first spurious
band at about 4.9~GHz.

As a final study, we now show the capabilities of the
hybrid manufacturing approach proposed in this work to
produce different types of transfer functions without 
changing the filter housing. The idea is to use the same housing to generate
different transfer functions by manufacturing new
dielectric pieces with the 3D printer.
To illustrate the basic idea, we now show the design of a new
air-hole filter having the same specifications as before,
except for the bandwidth which is increased to
$BW=175$~MHz.

\begin{figure}
\centering
\includegraphics[width=\columnwidth]{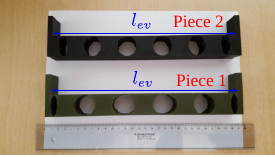}
\caption{Comparison of the dielectric pieces designed for the
air-holes filters having different bandwidths. Shown at the bottom
is the dielectric piece for the first filter with narrower bandwidth
(Piece~1). Shown at the top is the dielectric piece for the new
filter with wider bandwidth (Piece~2). Both pieces have the
same length: $l_{ev}=258.847$~mm.}
\label{fig:photo_airholes_different_bw}
\end{figure}

For the design of this new filter, an important geometrical
restriction must be taken into account, namely, that 
the total length ($l_{ev}$) for this
second piece must be the same as the one of the 
first filter ($l_{ev}=258.847$~mm). In
Fig.~\ref{fig:photo_airholes_different_bw} we show the dielectric piece
of the previous design (Piece~1), and we compare it with
the dielectric piece of the new design with increased
bandwidth (Piece~2).

As we can see, the same
length ($l_{ev}$) is preserved in both designs. The new
dielectric piece (shown on top), however, has smaller air holes to allow
for larger couplings. The smaller holes also increase the loading effects
of the couplings into adjacent resonators. Consequently, the
resonators become physically shorter, thus leading to air holes
which are placed closer as compared to the first design
(shown at the bottom). It is important to note that the total length
($l_{ev}$) can be made equal in both designs because of the
additional degree of freedom provided by the length
$l_1$ shown in Fig.~\ref{fig:evanescent_sketches}(a)
(distance between the step discontinuity and the first/last
air-hole).

The design procedure for the filter with increased bandwidth 
is the same as previously described.
The only difference is that new coupling and
$Q_{ext}$ curves
(similar to Fig.~\ref{fig:mij_curves} and Fig.~\ref{fig:qext_curves})
are obtained using (\ref{eqn:Mij}) and
(\ref{eq:qext}) with the new bandwidth ($BW$) specifications. The final
dimensions for this filter, as defined in 
Fig.~\ref{fig:evanescent_sketches}(a), are collected
in Table~\ref{tab:evanescente_aire_bw2}.
\begin{table}
\centering
\caption{Final dimensions of the designed air-holes filter
according to Fig.~\ref{fig:evanescent_sketches}(a),
for $BW = 175\,$MHz.}
\begin{tabular}{c|c||c|c||c|c} \hline
\textit{Var.} & Value (mm) & \textit{Var.} & Value (mm) &
\textit{Var.} & Value(mm)\\
\hline
$l_{port}$ & $10$ & $l_{1}$ & $8.25$ & $r_1$ & $6.752$ \\
$l_{d}$ & $7$ & $l_{2}$ & $22.059$ & $r_2$ & $11.702$ \\
$l_{ev}$& $258.847$ & $l_{3}$ & $23.255$ & $r_3$ & $13.623$ \\
$l_{tot}$& $292.847$ & $l_{4}$ & $23.411$ & $r_x$ & $14$ \\ \hline
\end{tabular}
\label{tab:evanescente_aire_bw2}
\end{table}

\begin{figure}
\centering
\includegraphics[width=\columnwidth]{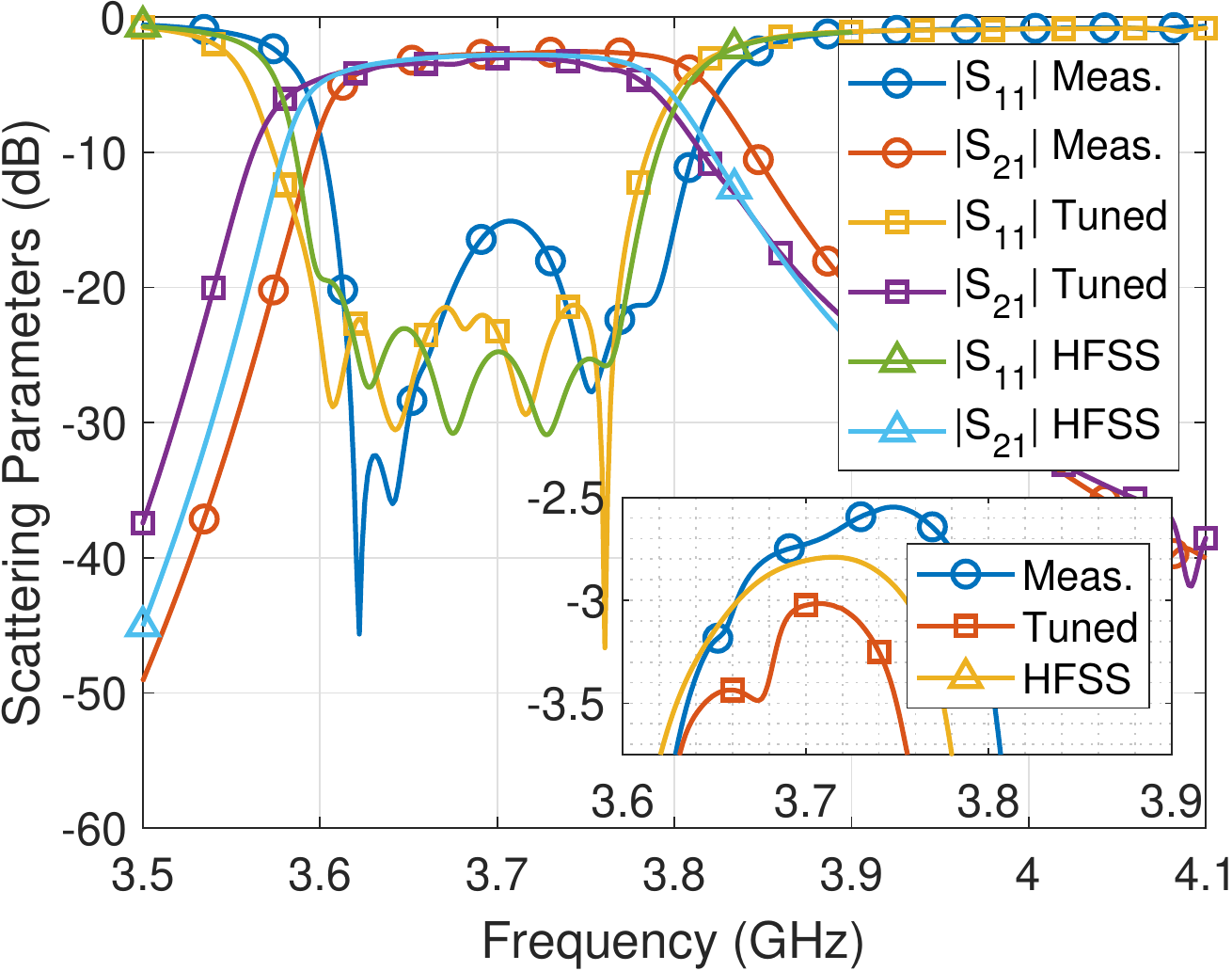}
\caption{Response of the air-holes filter designed with wider bandwidth
($BW=175$~MHz). Measured results directly after manufacturing
(circle symbols) are compared to results obtained after
tuning (square symbols), and with simulated results
obtained with the software tool HFSS (triangle symbols).}
\label{fig:airhole_measurements_175MHz}
\end{figure}

The new filter is obtained by replacing the 
dielectric Piece~1 with the dielectric Piece~2
inside the same metallic housing
(shown in Fig.~\ref{fig:photo_air_post_filter}).
The measured response obtained directly after manufacturing
is shown in Fig.~\ref{fig:airhole_measurements_175MHz}
(circle symbols).

In general, the performance obtained
is quite good in terms of center frequency and bandwidth
($f_c=3.71$~GHz, $BW=180$~MHz). The minimum insertion loss
inside the bandwidth is now lower due to the larger
bandwidth ($IL=2.5$~dB). In addition,
the return losses are also
good, being better than $RL=15$~dB inside the useful bandwidth.
In order to improve the return loss performance of the filter,
we have also used similar tuning screws as previously described
(see Fig.~\ref{fig:photo_air_post_screws}).
The response obtained after tuning is also shown in
Fig.~\ref{fig:airhole_measurements_175MHz} (square symbols).
As we can see, after the tuning process we have a 
passband with a center frequency of $f_c=3.687$~GHz, and
a bandwidth of $BW=172$~MHz. The measured results also show
an almost equiripple response with return losses better
than $RL=21$~dB. These improvements in the passband again
come at the expense of an increased insertion loss.
The minimum insertion loss within the passband 
is $IL=3$~dB. For comparison, the simulated response obtained
with the software tool HFSS is also included in
Fig.~\ref{fig:airhole_measurements_175MHz} (triangular symbols).
For this simulation the losses are characterized using the same
parameters as described in the previous examples.
As we can see, the losses predicted by the software
tool are very similar to those measured directly after
manufacturing, while the tuned prototype shows increased
losses due to the effects of the tuning screws.

\begin{figure}
\centering
\includegraphics[width=\columnwidth]{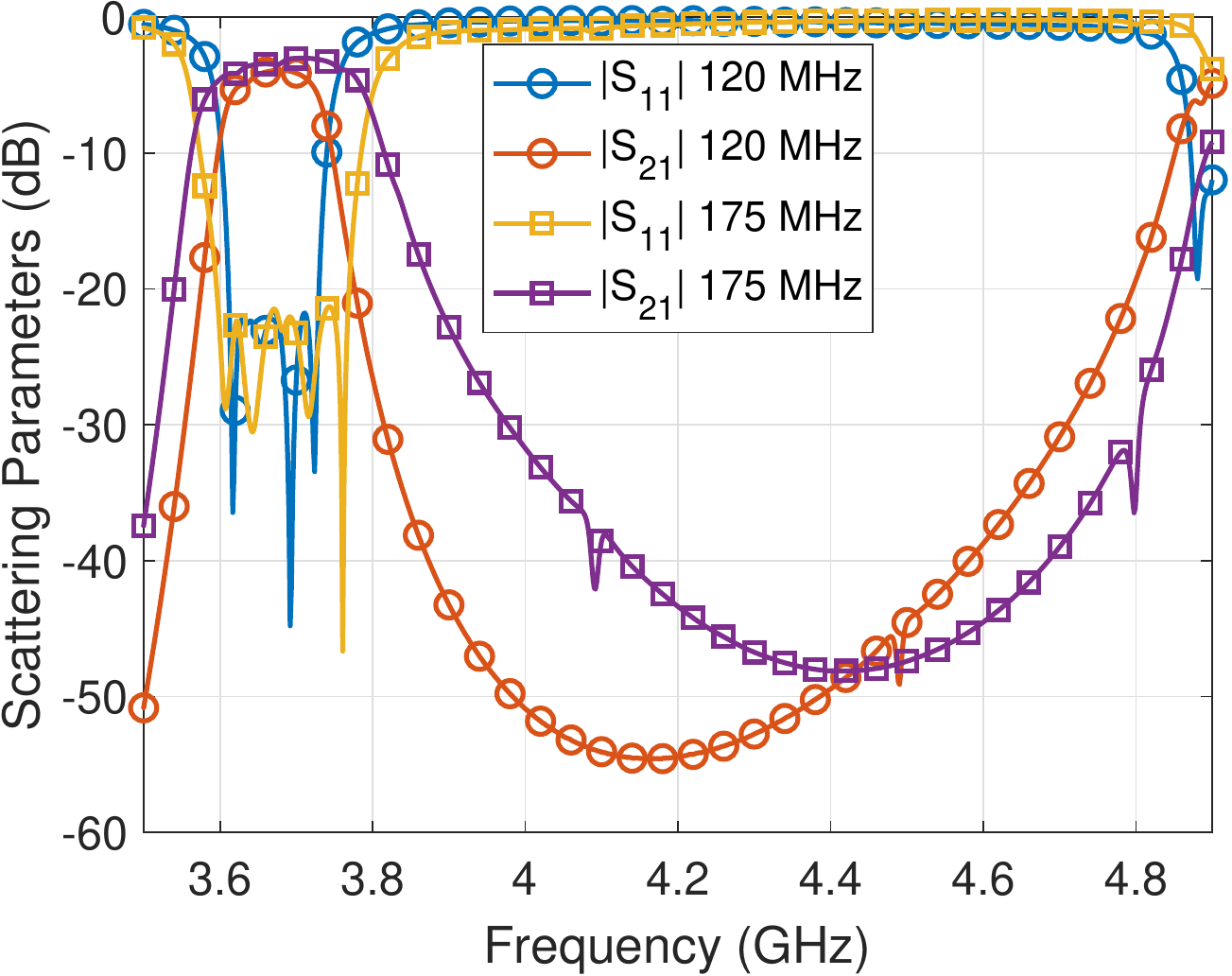}
\caption{Measured S-parameter responses of the two air-holes filters 
        with different bandwidths. Responses are shown in a large
	frequency span to monitor the out of band performances.}
\label{fig:airhole_two_bw}
\end{figure}

A final comparison between the two air-holes filters with 
different bandwidths
is presented in Fig.~\ref{fig:airhole_two_bw}. 
The measured responses after tuning have been selected
for this comparison. We can clearly see the differences in bandwidth
between the two filters. In both cases the return loss performance is
very satisfactory (better than $RL=20$~dB).
The filter responses are shown in a large frequency span, 
so that the SFR performances can be observed.
As we can clearly see, the first spurious band
again appears at $f_{sp} \approx 4.8$~GHz for both filters, 
giving $SFR \approx 1$~GHz.

Overall, the results presented in this paper clearly
show that the hybrid CNC milling and 3D printing manufacturing 
concept that we propose can be used
to obtain flexible and low cost microwave filters with 
acceptable electrical performances. The only drawback is to have
slightly higher values of insertion
losses. However, we expect that 
this drawback may be eliminated once better quality 3D printing 
materials become commercially available.  
\section{Conclusions}
The main aim of this work is to show experimentally 
that the use of 3D-printed ABSplus as dielectric material 
for the design of evanescent bandpass filters at C-band is indeed viable.
For this purpose,
two filter structures using dielectrics and
evanescent waveguide housings have been designed, 
manufactured and successfully measured. Both filters are based 
on an in-line topology, and combine
low cost additive manufactured parts for the dielectrics with
milled metallic parts for the external housing.

The first filter, based on air-holes, exhibits larger ($Q_U$)
value as compared to a solution based on
dielectric posts. However, the second option leads to more
compact structures, and to the possibility of achieving
higher bandwidths. The capability of implementing different
transfer functions, by 3D printing several different dielectric pieces
which are then integrated inside the same housing, has also been demonstrated.
The measured electrical performances
are satisfactory, but with slightly higher insertion losses due to the
electrical properties of the employed ABSplus plastic.
It is expected that improvements in insertion loss 
can be achieved as new 3D printing materials, with 
improved electrical properties, become commercially available.

From the results of this work, we conclude that
the hybrid CNC milling and 3D printing 
manufacturing approach demonstrated in this paper is, indeed,
a very valuable industrial tool for the rapid and flexible 
prototyping of practical microwave filters.   


\ifCLASSOPTIONcaptionsoff
  \newpage
\fi



%

\bibliographystyle{IEEEtran}
\bibliography{bibFile_paper}

\begin{thebibliography}{10}
\providecommand{\url}[1]{#1}
\csname url@samestyle\endcsname
\providecommand{\newblock}{\relax}
\providecommand{\bibinfo}[2]{#2}
\providecommand{\BIBentrySTDinterwordspacing}{\spaceskip=0pt\relax}
\providecommand{\BIBentryALTinterwordstretchfactor}{4}
\providecommand{\BIBentryALTinterwordspacing}{\spaceskip=\fontdimen2\font plus
\BIBentryALTinterwordstretchfactor\fontdimen3\font minus
  \fontdimen4\font\relax}
\providecommand{\BIBforeignlanguage}[2]{{%
\expandafter\ifx\csname l@#1\endcsname\relax
\typeout{** WARNING: IEEEtran.bst: No hyphenation pattern has been}%
\typeout{** loaded for the language `#1'. Using the pattern for}%
\typeout{** the default language instead.}%
\else
\language=\csname l@#1\endcsname
\fi
#2}}
\providecommand{\BIBdecl}{\relax}
\BIBdecl

\bibitem{boria07}
V.~Boria and B.~Gimeno, ``Waveguide filters for satellites,'' \emph{IEEE
  Microw. Mag.}, pp. 60--70, October 2007.

\bibitem{cameron-book}
R.~J. Cameron, C.~M. Kudsia, and R.~R. Mansour, \emph{Microwave Filters for
  Communication Systems: Fundamentals, Design and Applications}.\hskip 1em plus
  0.5em minus 0.4em\relax John Wiley, Sons; Inc. Hoboken, New Jersey, 2018, 2nd
  ed.

\bibitem{shang17}
X.~Shang, M.~Lancaster, and Y.-L. Dong, ``{W-band} waveguide filter based on
  large {TM120} resonators to ease {CNC} milling,'' \emph{Electron. Lett.},
  vol.~57, no.~7, pp. 488--490, March 2017.

\bibitem{dauria15}
M.~D'Auria, W.~J. Otter, B.~T.~W. Gillatt, C.~Long-Collins, N.~M. Ridler, and
  S.~Lucyszyn, ``{3-D} printed metal-pipe rectangular waveguides,'' \emph{IEEE
  Trans. Compon. Packag. Manuf. Technol.}, vol.~5, no.~9, pp. 1339--1349,
  September 2015.

\bibitem{peverini17}
O.~A. Peverini, M.~Lumia, F.~Calignano, G.~Addamo, M.~Lorusso, E.~P. Ambrosio,
  D.~Manfredi, and G.~Virone, ``Selective laser melting manufacturing of
  microwave waveguide devices,'' \emph{IEEE Proc.}, vol. 105, no.~4, pp.
  620--631, April 2017.

\bibitem{calignano17}
F.~Calignano, D.~Manfredi, E.~P. Ambrosio, S.~Biamino, M.~Lombardi, E.~Atzeni,
  A.~Salmi, P.~Minetola, L.~Iuliano, and P.~Fino, ``Overview on additive
  manufacturing technologies,'' \emph{IEEE Proc.}, vol. 105, no.~4, pp.
  593--612, April 2017.

\bibitem{montejo15}
J.~R. Montejo-Garai, I.~O. Saracho-Pantoja, C.~A. Leal-Sevillano, J.~A.
  Ruiz-Cruz, and J.~M. Rebollar, ``Design of microwave waveguide devices for
  space and ground application implemented by additive manufacturing,'' in
  \emph{2015 International Conference on Electromagnetics in Advanced
  Applications (ICEAA)}, Sep. 2015, pp. 325--328.

\bibitem{khan17}
S.~Khan, N.~Vahabisani, and M.~Daneshmand, ``A fully {3-D} printed waveguide
  and its application as microfluidically controlled waveguide switch,''
  \emph{IEEE Trans. Compon. Packag. Manuf. Technol.}, vol.~7, no.~1, pp.
  70--80, January 2017.

\bibitem{chan18}
K.~Y. Chan, R.~Ramer, and R.~Sorrentino, ``Low-cost ku-band waveguide devices
  using {3-D} printing and liquid metal filling,'' \emph{IEEE Trans. Microw.
  Theory Techn.}, vol.~66, no.~9, pp. 3993--4001, September 2018.

\bibitem{lorente09}
J.~A. Lorente, M.~M. Mendoza, A.~Z. Petersson, L.~Pambaguian, A.~A. Melcon, and
  C.~Ernst, ``Single part microwave filters made from selective laser
  melting,'' in \emph{Microwave Conference, 2009. EuMC 2009. European}, Sep.
  2009, pp. 1421--1424.

\bibitem{Guo16}
C.~Guo, X.~Shang, J.~Li, F.~Zhang, M.~J. Lancaster, and J.~Xu, ``A lightweight
  {3-D} printed {X-band} bandpass filter based on spherical dual-mode
  resonators,'' \emph{IEEE Microwave and Wireless Components Letters}, vol.~26,
  no.~8, pp. 568--570, Aug 2016.

\bibitem{guo15}
C.~Guo, X.~Shang, M.~J. Lancaster, and J.~Xu, ``A {3-D} printed lightweight
  {X-band} waveguide filter based on spherical resonators,'' \emph{IEEE
  Microwave and Wireless Components Letters}, vol.~25, no.~7, pp. 442--444,
  July 2015.

\bibitem{booth17}
P.~Booth and E.~V. Lluch, ``Enhancing the performance of waveguide filters
  using additive manufacturing,'' \emph{IEEE Proc.}, vol. 105, no.~4, pp.
  613--619, April 2017.

\bibitem{salonitis03}
K.~Salonitis, G.~Tsoukantas, P.~Stavropoulos, and A.~Stournaras, \emph{Virtual
  Modelling and Rapid Manufacturing - Advanced Research in Virtual and Rapid
  Prototyping}.\hskip 1em plus 0.5em minus 0.4em\relax CRC Press, Jan. 2003.

\bibitem{ngoc17}
N.-H. Tran, V.-N. Nguyen, A.-V. Ngo, and V.~Nguyen, ``Study on the effect of
  fused deposition modeling {(FDM)} process parameters on the printed part
  quality,'' \emph{Int. Journal of Engineering Research and Application},
  vol.~7, no.~12, pp. 71--77, December 2017.

\bibitem{carceller17}
C.~{Carceller}, F.~{Gentili}, D.~{Reichartzeder}, W.~{Bösch}, and
  M.~{Schwentenwein}, ``Practical considerations in the design of monoblock
  {TM} dielectric resonator filters with additive manufacturing,'' in
  \emph{2017 International Conference on Electromagnetics in Advanced
  Applications (ICEAA)}, Sep. 2017, pp. 364--367.

\bibitem{marchives14}
Y.~Marchives, N.~Delhote, S.~Verdeyme, and P.~M. Iglesias, ``Wide-band
  dielectric filter at {C-band} manufactured by stereolithography,'' in
  \emph{Microwave Conference (EuMC), 2014 44th European}, Oct. 2014, pp.
  187--190.

\bibitem{perigaud18}
A.~{Perigaud}, O.~{Tantot}, N.~{Delhote}, S.~{Verdeyme}, S.~{Bila}, and
  D.~{Baillargeat}, ``Bandpass filter based on skeleton-like monobloc
  dielectric pucks made by additive manufacturing,'' in \emph{2018 48th
  European Microwave Conference (EuMC)}, Sep. 2018, pp. 296--299.

\bibitem{ABSplus}
\BIBentryALTinterwordspacing
Stratasys. (2016, August) Absplus datasheet. [Online]. Available:
  \url{http://usglobalimages.stratasys.com/Main/Files/Material\_Spec\_Sheets/MSS\_FDM\_ABSplusP430.pdf}
\BIBentrySTDinterwordspacing

\bibitem{pons18}
A.~Pons-Abenza, A.~Alvarez-Melcon, F.~D. Quesada-Pereira, and
  L.~Arche-Andradas, ``Frequency correction design technique for additive
  manufactured cavity filters,'' in \emph{2018 48th European Microwave
  Conference (EuMC)}, Sep. 2018, pp. 288--291.

\bibitem{matthaei80}
G.~Matthaei, L.~Young, and E.~Yones, \emph{Microwave Filters, Impedance
  Matching Networks, and Coupling Structures}.\hskip 1em plus 0.5em minus
  0.4em\relax Boston, Massachusetts, USA: Artech House, 1980.

\bibitem{hfss}
ANSYS, ``{HFSS} version 19.2,'' https://www.ansys.com/products/ electronics/
  ansys-electronics-desktop, 2019.

\bibitem{3Dprinter}
\BIBentryALTinterwordspacing
Stratasys. (2017, May) Dimension {1200es} {3D} printer. [Online]. Available:
  \url{http://www.stratasys.com/es/impresoras-3d/design-series/dimension-1200es}
\BIBentrySTDinterwordspacing

\end{thebibliography}

%








\end{document}